\RequirePackage{fix-cm}
\documentclass[smallextended]{svjour3}       
\smartqed  
\usepackage{graphicx}
\usepackage{mathptmx}      
\usepackage[letterpaper,margin=1.25in]{geometry}
\usepackage{amsmath,amssymb}
\usepackage{bm}
\usepackage{color}
\usepackage{dsfont}
\usepackage{datetime}
\usepackage{graphicx}
\usepackage{algorithm}
\usepackage{verbatim}
\usepackage[usestackEOL]{stackengine}
\usepackage{tikz-cd}
\usepackage{wrapfig}

\everymath{\displaystyle} 

\journalname{Letters in Spatial and Resource Sciences}

\newcommand{\JELclassname}{{\bfseries JEL Subject Classification}\enspace}
\newcommand{\JELclass}[1]{\par\addvspace\medskipamount{\rightskip=0pt plus1cm
\def\and{\ifhmode\unskip\nobreak\fi\ $\cdot$
}\noindent\JELclassname\ignorespaces#1\par}}

\begin{document}

\title{Optimal spatial-dynamic management to minimize the damages
  caused by aquatic invasive species\thanks{The work of Zipp and
  Y. Wu was partially supported by the Department of Agricultural
  Economics, Sociology, and Education at Penn State. The work of
  Zikatanov was partially supported by NSF grants DMS-1720114 and
  DMS-1819157.}}


\titlerunning{Optimal management to minimize the damages of AIS}    

\author{Katherine Y. Zipp \and Yangqingxiang Wu \and Kaiyi Wu \and Ludmil T. Zikatanov}

\authorrunning{K.~Zipp, Y.~Wu, K.~Wu, L.~T.~Zikatanov} 

\institute{Katherine Y. Zipp \at
              Department of Agricultural Economics, Sociology and Education, Penn State, University Park, PA 16802 \\
              \email{kyz1@psu.edu}           
           \and
           Yangqingxiang Wu \at
             Department of Mathematics, Penn State, University Park, PA 16802\\
              \email{yzw137@psu.edu}           
           \and
          Kaiyi Wu \at
             Department of Mathematics, Tufts University, Medford, MA 02155\\
              \email{Kaiyi.Wu@tufts.edu}           
            \and
            Ludmil T. Zikatanov \at
             Department of Mathematics, Penn State, University Park, PA 16802\\
              \email{ludmil@psu.edu}           
}

\date{Received: date / Accepted: date}

\maketitle

\begin{abstract}
Invasive species have been recognized as a leading threat to
biodiversity. In particular, lakes are especially affected by species
invasions because they are closed systems sensitive to
disruption. Accurately controlling the spread of invasive species requires solving
a complex spatial-dynamic optimization problem. In this work we propose a
novel framework for determining the optimal management strategy to maximize the value of a lake system net of damages from invasive species, including an endogenous diffusion mechanism for the spread of invasive species through boaters' trips between lakes. The proposed method includes a combined global iterative process
which determines the optimal number of trips to each lake in each
season and the spatial-dynamic optimal boat ramp fee.

\JELclass{Q20 \and Q50 \and Q57}
\end{abstract}

\section{Introduction}

Globally, invasive species have long been recognized as a leading threat to biodiversity \cite{Wilcove1998,Sala2000}. Lakes are especially affected by species invasions because they are closed systems sensitive to disruption \cite{Sala2000,Moorhouse2015}. As a result, controlling the spread of aquatic invasive species (AIS) has been a major management effort for the past two decades. Further complicating AIS management, AIS are largely spread inadvertently through the movement of recreational boaters from lake to lake \cite{Rothlisberger2010}. Therefore, management is tasked with maximizing the value of a lake system net of AIS damages by changing boating behavior and thus the spread of AIS.

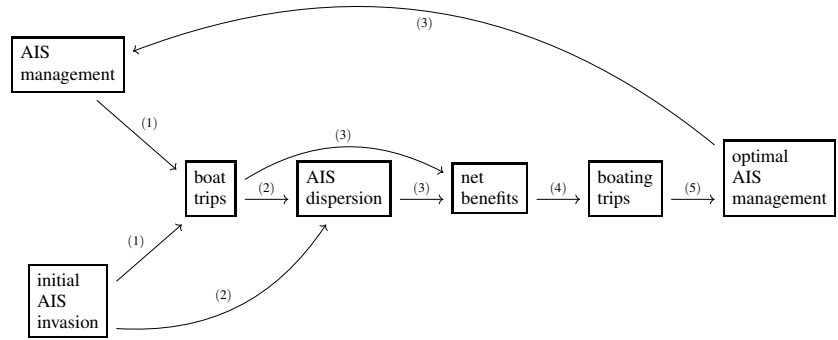
\begin{figure}[!htb]
  \begin{center}
  \begin{scriptsize}
\begin{tikzcd}
\framebox{\Longstack[l]{AIS\\management}}
\arrow{dr}{(1)}  & & & & & \\
  & \framebox{\Longstack[l]{boat\\ trips}} \arrow{r}{(2)}\arrow[bend left]{rr}{(3)} & \framebox{\Longstack[l]{AIS\\ dispersion}}\arrow{r}{(3)} & \framebox{\Longstack[l]{net\\ benefits}} \arrow{r}{(4)} &\framebox{\Longstack[l]{boating\\ trips}} \arrow{r}{(5)} &\framebox{\Longstack[l]{optimal\\AIS\\management}}\arrow[bend right]{ulllll}{(3)}\\
\framebox{\Longstack[l]{initial\\ AIS\\ invasion}}
\arrow{ur}{(1)}\arrow[bend right]{urr}{(2)}  & & & & &  
\end{tikzcd}
\end{scriptsize}
  \caption{Diagram of model framework \label{fig:model}}
  \end{center}
\end{figure}

In this article, we develop a novel spatial-dynamic framework to determine the optimal policy to induce the optimal number of trips to each lake in a system across many seasons to maximize the net benefits of the lake system taking into consideration the damages from the spread of invasive species. 
This policy must be heterogeneous across space and time. Spatially, it must depend on boating patterns (which depend on the attractiveness of lakes, substitutability across lakes, the presence of invasive species) and the ecological suitability of a lake for invasion.  Temporally, it must take into account the time dependent distribution of AIS across the system. AIS spread is a vicious cycle, as an increase in invaded lakes provides more opportunity for spread to uninvaded lakes. The pathways for dispersal of AIS depend on boating behavior, which in turn depends on lake management and the presence or absence of AIS in each lake. This leads to a complicated feedback loop where the optimal policy depends on the dispersal of AIS and the number of trips to each lake, the dispersal of AIS depends on boating behavior, and boating behavior depends on policy and the dispersal of AIS (see Figure \ref{fig:model}). The dispersal of AIS and boating behavior both depend on the presence or absence of AIS in every lake in the system. Therefore, the optimal policy must be globally coordinated across the system \cite{Epanchin-niell2010}.

There are five major components to our optimal spatial-dynamic AIS management framework. 

\begin{enumerate}
	\item \textbf{Model of recreational boating decisions}. The dispersal of AIS depends on boating decisions, therefore, we need a model of how boaters choose where to boat. We rely on a standard economic model known as a random utility model (RUM) in which boaters maximize their utility by choosing where to boat. 
	\item \textbf{Model of AIS dispersal}. The dispersal of AIS requires (i) boaters to visit an invaded lake, (ii) inadvertently transport the invasive species out of the invaded lake, (iii) then visit an uninvaded lake while the invaded species is still alive, and finally (iv) the invasive species must become established in a suitable lake \cite{Zipp2018}. Given the dispersal of AIS, we model the probability that a lake becomes invaded given that it is not already invaded as a hazard model. 
	\item \textbf{Net benefits to recreational boaters}. Lakes provide benefits to boaters. Our RUM allows us to calculate the welfare benefits of boating. The net benefits of the lake system are these benefits of boating minus the damages caused by AIS. We consider both the damages from AIS to both shoreline property owners \cite{Horsch2009,Provencher2012} and boaters \cite{Lewis2015}. 
	\item \textbf{Optimal number of boating trips to each lake}. We find the optimal number of boating trips to each lake in each season to maximize net benefits (step 3) subject to AIS dispersal (step 2) and the model of boating decisions (step 1). 
	\item \textbf{Optimal policy}. Finally, we find the optimal policy that leads boaters to choose the socially optimal number of trips. 
\end{enumerate}

This framework makes three major contributions to the literature. First, we develop a method that has not been used in the emerging literature on optimal spatial-dynamic policy management. 

 \cite{Epanchin-niell2010,Brock2008,Brock2010,Sanchirico2010,Eiswerth2002,Hof1998,Horan2005,Leung2002}. Previous literature has focused on dynamic optimization procedures that are plagued by the curse of dimensionality. Our framework does not suffer this problem and does not require large computing power to implement. Second, we incorporate an endogenous dispersal method. The literature on optimal spatial-dynamic bioeconomic policies typically models the dispersal of species as radially dispersing from an original location. Our model on AIS allows the dispersal mechanism to depend on the movement of boaters and thus be endogenous.  Third, we have a fully-coupled natural human system where boating decisions depend on the status of invasions and the dispersal of invasions depends on boating decisions.

\section{Problem Formation}

The spatial-dynamic optimal policy can be determined through an iterative process with five major components (see Figure \ref{fig:model}), which will be explained in detail. 

\subsection{Model of recreational boating decisions}

We model boating decisions as a repeated random utility travel cost model in which boaters $n\in \{1, \ldots, N\}$ maximize their utility
$U_{isnt}$ on day $t\in \{1, \ldots, T\}$ in season $s \in
\{1, \ldots, S\}$ by either (a) visiting lake $i \in \{2, \ldots, I\}$; or
(b) choosing not to go boating (denoted by $i=1$). Let utility be defined as

\begin{equation}\label{utility}
  \begin{array}{l}
    \displaystyle
    U_{isnt} = v_{isnt} + \varepsilon_{isnt}, \quad \text{where}\\
    v_{isnt} = Z_{isnt} + \alpha (M_n - \tau_{i,s}) - \xi x_{i(s-1)} ,  
    \end{array}
\end{equation}
$Z_{isnt}=\sum\nolimits_{m}\beta_m z_{isnt,m}$ represents $m$ boater, lake, day, or season
attributes ($z_{isnt,m}$) that influence decision to go boating or stay home with parameters $\beta_m$ and remains constant in our model, $\alpha$ is marginal utility of income, $M_n$ is the income of boater $n$, $\tau_{is}$ is the boat ramp fee, $\xi$ is the effect of AIS on boater utility, $x_{i(s-1)}$ is the status of AIS lake $i$ at the end of the
previous season $(s-1)$\footnote{We assume that the invasion status is
  updated at the end of the season such that boaters' trip decisions
  depend on $x_{s-1}$ and at the end of the season when the invasion
  status is updated $x_s$ depends on the boating decisions in season
  $s$.}, and $\varepsilon_{isnt}$ are identically independently distributed
Type I Extreme Value random variables with variance $\pi^2/6$. Under this assumption of error distribution, the probability that boater $n$ chooses an
alternative $i \in \{1, \ldots I\}$ on day $t$ in season $s$ is given by the conditional logit model \cite{Train2009} 
\begin{equation}\label{P}
P_{isnt} = \frac{\exp(v_{isnt})}{\sum_{j=1}^{I}\exp(v_{jsnt})}. \nonumber
\end{equation}
We set the total number of trips taken to lake $i \in \{2,\ldots,I\}$ (or days spent at home in the case of $i=1$)
in season $s$ to be \(b_{is} = \sum_{n=1}^N \sum_{t=1}^T P_{isnt}\) (note that
$\sum_{i=1}^I b_{is} =
NT$ by definition). Clearly, $b_{is}$
is a function of the boat ramp fees $\{\tau_{js}\}_{j=2}^I$ and the
invasion statuses $\{x_{j(s-1)}\}_{j=2}^I$.

\subsection{Model of AIS dispersal}

AIS are largely spread inadvertently through the movement of boaters across lakes \cite{Chivers2012,Rothlisberger2010}. We define the status of the invasion in lake $i$ in season $s$, denoted by $x_{is}$, as the probability that lake $i$ is invaded at the end of season $s$. We assume that invasions are irreversible so that once a lake becomes invaded it remains invaded for all remaining seasons. Let $\widetilde{x}_{is}$ be the probability that lake $i$ is invaded at the end of season $s$ conditional on being uninvaded at the end of season $s-1$. We have the following relations.
\begin{equation}\label{x}
  x_{i(s+1)}=1-\prod_{m=1}^{s}(1-\tilde{x}_{im}),\quad \mbox{or equivalently},
  \quad x_{i,s+1}= x_{is}+(1-x_{is})\tilde{x}_{i(s+1)}
\end{equation}
We assume that $x_{i,s=1}=:x_{i1}$ is given.
We use the following approximate hazard
model:
\begin{equation}\label{e:no-approx}
\widetilde{x}_{is} = \mathfrak{s}(K_{is}) = \begin{cases}
0, & K_{is}< -a,\\
\frac{K_{is}+a}{2a}, & K_{is} \in [-a,a],\\
1, & K_{is}> a,
\end{cases}
\end{equation}
which approximates the sigmoid function\footnote{We take $a=2.824153$
  which gives an approximates the sigmoid with $\ell_\infty$ error at most $0.056075$.}. Here, $K_{is}$ is
defined as
\[
K_{is} = \gamma L_{is}+ \lambda_{is}
\] 
where $L_{is}$ is the expected number of trips carrying AIS to lake $i$ in season $s$ (referred to as propagule pressure),  $\lambda_{is}$ is a measure of the suitability of lake $i$ to host AIS, and $(\gamma,\xi_l$) are estimated parameters related to the effects of propagule pressure and lake suitability on the probability of invasion. Propagule pressure requires (i) boaters to visit an invaded lake, (ii) inadvertently transport the invasive species out of the invaded lake, and (iii) then visit an uninvaded lake while the invaded species is still alive. Recent evidence suggests that the majority of AIS remains alive for approximately one day \cite{Bruckerhoff2014}. It is not possible to spread the invasive species from lake $i$ to itself, so we are only concerned with trips between different lakes. Therefore, the expected probability that a boater visits any invaded lake $j\neq i$ on the previous day is $\sum\nolimits_{j\neq i}P_{jsn(t-1)}x_{j(s-1)}$. Let $\mu$ be the probability of a boat leaving an invaded lake with AIS (this is a constant independent of $n$,$i$,$s$,$t$). Therefore, the propagule pressure is
\[
L_{is} = \sum_{t=2}^T \sum_{n=1}^N
\underbrace{\left( \mu \sum_{j\neq i} 
P_{jsn(t-1)}	x_{j(s-1)}\right) }_{Q_{isnt}}P_{isnt}
\]
where $Q_{isnt}$ is the expected number of trips that boater $n$ makes to lake $i$ on day $t$ in season $s$ carrying an invasive species from any lake $j\neq i$.

\subsection{Net benefits}

Lakes provide benefits to boaters who receive utility from boating instead of staying home. Let $w_{snt} = \max \left\{v_{1snt}, \ldots, v_{Isnt}\right\} $. Without any boating, boaters just receive utility from staying home $w_{snt} = v_{1snt}$. Therefore, the value of having the
option to boat at all lakes in the system $i \in {2, \ldots, I}$ is
worth $\mathcal{L}^{s}_{snt}$ per boater per trip per season:
\begin{equation}\label{value of lake system}
\begin{array}{l@{{}={}}l}
\mathcal{L}^{s}_{snt} & -\frac{1}{|\alpha|}\left(\ln\left(\exp(v_{1snt})\right)-\ln\left(\sum_{i=1}^{I}\exp(v_{isnt})\right)\right) \\
& -\frac{1}{|\alpha|} \left(\ln\left( \frac{\exp(v_{1snt})}{\sum_{i=1}^{I}\exp(v_{isnt})} \right) \right) \\
& -\frac{1}{|\alpha|}\left( \ln \left(P_{1snt}\right)\right)  \geq 0,
\end{array}
\end{equation}
where $\alpha$ is the marginal utility of income and is used to
convert measures of utility into measures of dollars and $P_{1snt}\leq
1$ is the probability of boater $n$ choosing to stay home $i=1$ on day
$t$ in season $s$. The total value of the system to all boaters across all trips in season $s$ is the sum of $\mathcal{L}^{s}_{snt}$ over boaters $n$ and days $t$: 
\begin{equation}\label{wss}
\mathcal{W}^{s} _{s} = -\frac{1}{|\alpha|} \sum_{n,t} \ln(P_{1snt})\approx
-\frac{1}{|\alpha|}\sum_{n,t} (P_{1snt}-1) =
-\frac{1}{|\alpha|}\left(b_{1s} - NT\right) =
\frac{1}{|\alpha|} \sum_{i=2}^I b_{is}. 
\end{equation}

The net benefits of the lake system are these benefits $\mathcal{W}_s^s$ minus the damages caused by AIS. Aquatic invasive species (AIS) cause damage to both shoreline property owners \cite{Provencher2012,Horsch2009} and recreational users (e.g. boaters) \cite{Lewis2015}. We consider $h_i$ shoreline properties around lake $i$, assuming that shoreline properties are constant across seasons. We assume that shoreline property owners incur a constant annual welfare loss, $\mathcal{L}^{h}$, per shoreline property, $h_i$, for each invaded lake. Therefore, we define the total welfare loss in season $s$ from the spread of AIS to shoreline property owners as

\[
\mathcal{W}^{h}_s= \sum_{i=2}^I\mathcal{L}^{h} h_i x_{is}
\] where we use that $x_{1s}=0$ for all $s$.

Next, we define the welfare loss from the spread of AIS to boaters. We assume that boaters incur a constant welfare loss, $\mathcal{L}^{b}$, per trip $b_{is}$ to an invaded lake in season $s$. Therefore, we define the expected total welfare loss in season $s$ from the spread of AIS to recreational boaters as 

\[
\mathcal{W}^{b}_s = \sum_{i=2}^I\mathcal{L}^{b}b_{is}x_{is}. 
\]

Therefore, the net benefits of the entire lake system is the discounted sum of the benefits of the lake system, $\widetilde{\mathcal{W}_s^s}$ minus the damages from AIS, $\mathcal{W}^{h}_s + \mathcal{W}^{b}_s$:

\begin{equation}\label{damages}
\mathcal{F} = \sum_{s} \rho^s \left(\widetilde{\mathcal{W}_s^s} - \mathcal{W}^{h}_s - \mathcal{W}^{b}_s \right)
\end{equation} where $\rho^s = \left(\frac{1}{1+r}\right)^s$ is the discount factor and $r$ is the discount rate.

\subsection{Optimal number of boating trips to each lake}

With this net benefit function, $\mathcal{F}$, our objective is to find the optimal boat ramp fee
$\tau_{is}^*$ that maximizes the net benefit of the lake system subject to the dispersal function of AIS \eqref{xis2}. This is a difficult optimization problem to solve. We make two contributions that allow us to more easily solve this problem. First, we rewrite the objective function $\mathcal{F}(\bm{b}(\bm{x},\bm{\tau}),\bm{x}(\bm{b},\bm{\tau}))$ as a function 
of $\bm{b}$ in the following general form
\begin{equation}\label{quadform}
  \mathcal{F}(\widetilde{\bm{b}},
  \bm{b}) = \langle \mathbb{D}(\widetilde{\bm b})\bm{b},\bm{b} \rangle +
2 \langle\bm{f}(\widetilde{\bm b}),\bm{b}\rangle + \langle \bm{g}(\widetilde{\bm b},\bm{P}),\bm{1}\rangle. 
\end{equation}
The reason this is can be written as function of $\bm{b}$ is
because, as discussed after Algorithm~4, the boat ramp fee $\bm{\tau}$ in the
model can be determined by the number of trips $\bm b$, and the
probabilities of invasion $\bm x$ are also uniquely determined from
the taxes $\bm\tau$.  It is important to note that in general we would
like to maximize $\mathcal{F}(\cdot,\cdot)$ above when $\widetilde{\bm
  b}=\bm{b}$. Thus, for a fixed $\widetilde{\bm b}$ we have the
following maximization to find the optimal number of trips to each
lake $\bm b^{*}$ to maximize net benefits with three constraints:(i)
AIS dispersal follows \eqref{xis2}, (ii) the total number of trips has
to equal the number of trip occasions \eqref{constraints}, (iii) the
minimum number of trips boaters can take to lake $i$ is zero ~\eqref{constraints}. \\

\noindent\fbox{%
	\parbox{\textwidth}{%
		
		\begin{center}
			{\bf Optimization Problem}
		\end{center} 
		\begin{equation}
		  \mbox{Find}\quad \bm \tau_{*} :=\arg\max_{\bm{\tau}}
                  \mathcal{F}(\bm{b}(\bm{\tau}),\bm{P}(\bm{\tau}) =
		   \left\{
                         \langle\mathbb{D}(\bm b) \bm b,\bm b\rangle
			 + 2\langle\bm f(\bm b),\bm b\rangle  +
			 \langle g(\bm b,\bm{P}(\bm{\tau})),\bm 1\rangle
                          \right\},
		\end{equation}
          	Subject to the constraints:
		\begin{equation}\label{constraints}
                  \sum\nolimits_{i=1}^I b_{is} = N\times T, \quad
		  0 \leq b_{is}
		\end{equation}                
	        where $x_{is}$ is computed with given initial values
                $\{x_{i1}\}_{i=1}^I$ and
                \begin{equation}\label{xis2}
		x_{is} = x_{i,s-1} +  (1-x_{i,s-1})\mathfrak{s}(K_{is})
		\end{equation}
                for $i=2,\ldots,I$ and $s=2,\ldots,S$.
	}%
}

In what follows we will use iterative approximation of $\mathcal{F}$
by quadratic forms which can be easily optimized.
In such cases we evaluate all terms such as $\mathbb{D}(\bm{b})$,
$\bm{f}(\bm{b})$, 
$\bm{g}(\bm{b},\bm{P})$ at our previous iterates, denoted here by
$\widetilde{\bm{b}}$, $\widetilde{\bm{\tau}}$. We thus have
\[
\mathcal{F}(\widetilde{\bm{b}},\bm{b}):=
\langle\mathbb{D}(\widetilde{\bm b}) \bm b,\bm b\rangle + 2\langle\bm
f(\widetilde{\bm b}),\bm b\rangle + \langle g(\widetilde{\bm
  b},\bm{P}(\widetilde{\bm{\tau}})),\bm 1\rangle.
\]

To be more specific, let us introduce the quantities involved in
writing the explicit functional form of $\mathbb{D}$, $\bm{f}$, and
$\bm{g}$. We define:
\begin{eqnarray*}
	&&  m_{1,is} = \frac{1-x_{i,s-1}}{2a} w_{is}+\frac{1+x_{i,s-1}}{2},\quad 
	m_{2,is} = \frac{1-x_{i,s-1}}{2a}y_{is}A_s\\
	&&  c_{1,is} = L^bm_{1,is} + L^h h_i m_{2,is}, \quad
	c_{2,is} = L^b m_{2,is}.
\end{eqnarray*}
A direct calculation from the definition of $\mathcal{F}$ gives
\begin{eqnarray*}
&&	\mathbb{D}_{is,is}  = \begin{cases}
		- \rho^s c_{2,is}, & K_{is}\in [-a,a],\\
		0, & K_{is}\notin [-a,a],
	\end{cases},\quad 
	g_{is} = \begin{cases}
		-\rho^sL^h h_im_{1,is}, & K_{is}\in [-a,a],\\
		0, & K_{is}<-a,\\
		-\rho^s L^h h_i, & K_{is}>a,
	\end{cases}\\      
&&	f_{is}  =  \frac{1}{2}\begin{cases}
		\rho^s \left(c_{1,is}-\frac{1}{|\alpha|}\right), & K_{is}\in [-a,a],\\
		\frac{\rho^s}{|\alpha|}, & K_{is}<-a,\\
		\rho^s \left(L^b-\frac{1}{|\alpha|}\right), & K_{is}>a,
	\end{cases}
\end{eqnarray*}

\subsection{Optimal policy}
To compute optimal tax values we propose the following algorithms to
find the optimal boat ramp fees, $\tau^*_{is}$ for each lake $i$ in
season $s$, that map from the optimal number of trips to each lake,
$b_{*,is}$.

\begin{algorithm}[htb]\caption{Global iteration\label{global-iter}}
	\renewcommand{\labelenumi}{\arabic{enumi}.}
	\noindent	
	\begin{enumerate}	
		\item Input: given an initial state of invasions $x_{i,s=1}$ (a vector with $I$
		elements) and an initial guess of the boat ramp fee $\tau_{is}^{(0)}$ (an $(I\times S)$ matrix), we propose the following iterative procedure.
		\item[] {\bf For}  $k=1,2,\ldots$ until convergence:
		\begin{enumerate}
			\item Compute the spread of AIS
			$x_{is}^{(k)}$ for all $s\in (2,\ldots,S)$ from $\tau_{is}^{(k)}$ using Algorithm~\ref{alg:tau2x}.
			\item Find the $b_{is}^{*(k)}$ that maximizes
                          the net benefits via
                          Algorithm~\ref{alg:minb} for fixed $\bm P$
                          from the previous iteration.
			\item Find $\tau_{is}^{*(k)}$ from the
                          computed $b_{is}^{*(k)}$ by solving the
                          nonlinear system of equations relating these
                          quantities via Algorithm~\ref{alg:b2tau}.
		\end{enumerate}
		
		\item {\bf If} a convergence criteria is met, or iteration number is too large {\bf Then} Stop 
		
	\end{enumerate}
	
\end{algorithm}

We now rewrite $L_{is}$ as a linear function
of $b_{is}$ plus some ``noise''. To do this,
let us introduce $A_{is}$ -- the average of $Q_{isnt}$ with respect to $n$ and $t$ or the share of boater trips where the boater removes an invasive species from an invaded lake, namely. 
\[
A_{is}=\frac{1}{NT}\sum_{n}\sum_{t} Q_{isnt}=
\frac{1}{NT}\sum_n\sum_t\left[\sum_{j\neq i} 
\mu x_{j(s-1)}P_{jsn(t-1)}\right].
\]
Then for $L_{is}$ we have, 
\[
L_{is} = A_{is}b_{is} + \underbrace{\sum_t \sum_n  (Q_{isnt}-A_{is}) P_{isnt}}_{g_{is}}
\]
In the algorithms that follow, the second term in the right side above
is lagged, i.e. taken from the previous iteration.  As a result, e can
write $K_{is}$ as a linear function of $\bm{b}$:
\[
K_{is} = \gamma A_{is}b_{is} + \gamma g_{is}(\bm{b},\bm{P})+ \lambda_{is}.
\] 
where $A_{is}b_{is}$ is the average propagule pressure on lake $i$, i.e. the average number of trips to lake $i$ that carry an invasive species in season $s$, and $g_{is}$ is a measure of the deviance across trip occasions from the average propagule pressure.

The algorithms described below, follow the steps outlined in the Global Iteration Algorithm~\ref{global-iter}. 
\begin{algorithm}[htb]\caption{Compute $x_{is}$ given $\tau_{is}$\label{alg:tau2x}}
	\renewcommand{\labelenumi}{\arabic{enumi}.}
	\noindent
	\begin{enumerate}
		\item {\bf Input}: 
		a vector $x_{\cdot,1}=(x_{11},x_{21},\ldots,
		x_{I,1})^t \in \mathbb{R}^I$ (for $s=1$) and $\tau_{is}^{(k)}\in
		\mathbb{R}^{I\times S}$.
		
		\item {\bf Output}: $x_{\cdot,2:S+1}^{(k)}\in \mathbb{R}^{I\times S}$ and $b_{is}^{(k)}$.
		
		\textbf{For} $s=2,...,(S+1)$
		\begin{itemize}\setlength{\itemsep}{\normalbaselineskip}
			
			\item[]  $P_{isnt}^{(k)} = 	\frac{ \exp \left(\sum_m z_{isnt,m}\beta_m -\xi x_{i(s-1)}^{(k)} -\alpha\tau_{is}^{(k)} \right) }
			{\sum_{j}  \left\{ \exp \left(\sum_m z_{jsnt,m}\beta_m  -\xi x_{j(s-1)}^{(k)} - \alpha \tau_{js}^{(k)} \right)  \right\} }$ 
			\bigskip
		      \item[] $b_{is}^{(k)} = \sum_{n,t} P_{isnt}^{(k)}$,
                        \quad $Q_{isnt}^{(k)} = \sum_{j\neq i} \mu P_{jsn(t-1)}^{(k)}x_{j(s-1)}^{(k)} $
		      \item[] $A_{is}^{(k)} = \frac{1}{NT}\sum_{n,t} Q_{isnt}^{(k)}$, \quad
                        $g_{is}^{(k)} = \sum_{n,t} \left( Q_{isnt}^{(k)}-A_{is}^{(k)} \right) P_{isnt}^{(k)}$
                        \bigskip
		      \item[] $ d_{is}^{(k)} = \gamma_{is}g_{is}^{(k)}$,
                        \quad $ K_{is}^{(k)} = \gamma_{is}A_{is}^{(k)}b_{is}^{(k)} + d_{is}^{(k)} + \lambda_{is}$ 
			
                        Now, for a given $b_{is}$, we update $x_{is}$
			\begin{equation}\nonumber
			    x_{is}^{(k)} =
			    x_{i,s-1}^{(k)}+  \mathfrak{s}\left(K_{is}^{(k)}\right)
			\end{equation}
			
			
		\end{itemize}
	\end{enumerate}
\end{algorithm}

\begin{algorithm}[htb]\caption{Maximization (Find $\bm{b}$ given
    $\widetilde{\bm{b}}^{(k)}$ and $\bm{x}$))
		\label{alg:minb}}
	\renewcommand{\labelenumi}{\arabic{enumi}.}
	
	\begin{enumerate}
		\item {\bf Input}: $\bm{x}_{\cdot,2:S+1}\in \mathbb{R}^{I\times S}$, $\widetilde{\bm{b}}^{(k)}$.
		\item {\bf Output}: $\bm{b^{(k+1)}}\in\mathbb{R}^{I\times S}$.
	
		\item Set $\widetilde{\bm{b}}=\bm{b}^{(k)}$
                  \item Find the $\bm b^{(k+1)}$:    
		\(\bm b^{(k+1)}=\arg\max_{\bm{b}}
                \mathcal{F}(\bm{b}^{(k)},\bm{b})\),
                subject to the constraints:
                \begin{equation*}
                  0\le b_{is}\leq \upsilon_{is} := \sum_{n,t}
                  \frac{\exp\left(v_{isnt}(\cdot, \tau_{is}^{(k)})\right)}{\sum_{j=1}^{I}\exp\left(v_{jsnt}(\cdot, \tau^{(k)}_{js})\right)}.
                  \end{equation*}
	\end{enumerate}
\end{algorithm}
Notice that $\mathcal{F}(\widetilde{\bm b},\cdot)$ is a quadratic
functional for fixed $\widetilde{\bm b}$ with negative definite
diagonal matrix $\mathbb{D}$. The solution then to the optimization
problem in Algorithm ~\ref{alg:minb}, is given by (see
Deutsch~\cite[Theorem~4.1, p.43]{frank2001})
		\begin{equation}\label{bstar}
		b_{is}^{(k+1)} = \begin{cases}
		0, & [\mathbb{D}^{-1} \bm{f}]_{is}<0\\
		[\mathbb{D}^{-1}\bm{f}]_{is}, & 0\le [\mathbb{D}^{-1} \bm{f}]_{is}\le \upsilon_{is}\\
		u_{is}, & [\mathbb{D}^{-1} \bm{f}]_{is}>\upsilon_{is}.
		\end{cases}
		\end{equation}
                
\begin{algorithm}[htb]\caption{Computation of $\tau_{is}^{(k+1)}$
		from $b_{is}^{(k+1)}$ and $x_{is}^{(k)}$.\label{alg:b2tau}}
	\renewcommand{\labelenumi}{\arabic{enumi}.}
	\noindent
	\begin{enumerate}
	\item {\bf Input:} $\bm b = \bm b^{(k+1)}$ from Algorithm~\ref{alg:minb} and $\bm x^{(k)}$ from Algorithm~\ref{alg:tau2x}.   
		\item {\bf Output:} $\bm \tau^{(k+1)}$ which solves
		  $\mathcal{R}(\bm \tau^{(k+1)}) =\bm b^{(k+1)}$,
                  where, for $i\ge 2$, \\
		$\mathcal{R}(\tau_{is}^{(k+1)})=\sum_n\sum_t \frac{
			\Theta_{isnt}^{(k)}\eta_{is}^{(k)}} {\left[
			\sum_{j}\Theta_{jsnt}^{(k)}\eta_{js}^{(k)}\right] }$ and
		$\Theta_{isnt}^{(k)}=\exp\left( \widetilde{Z}_{isnt}^{(k)}\right),
		  \eta_{is}^{(k)}=\exp(-\alpha\tau_{is}^{(k)})$, and,
		 \[
		\widetilde{Z}_{isnt}^{(k)} = \sum_m z_{isnt,m}\beta_m - \zeta x_{i(s-1)}^{(k)} .
		\]		
                  \\
		
		This nonlinear system is solved by the following iterative method:
		\begin{itemize}
			
			\item[] {\bf Set}
			$\eta_{1s} = 1$ (or $\tau_{1s}^{(k)} = 0$), $S=2,\ldots, S+1$. This reduces the number of unknowns, but number of equations is also reduced as
			$b_{1s} = NT - \sum_{i\ge 2} b_{is}$.
			\item[] {\bf Set} an initial guess $\bm{\tau^{(0)}_{is}}$ (which gives us $\eta_{is}^{(0)}$). 
			
			\item[] {\bf For} $q=1,2,\ldots$ until convergence:
			
			\item[] {\bf For}  $s=2,\ldots, S+1$,$i=2,..., I$ solve for $\eta^{(q)}_{is}$ 
			\begin{equation}\label{eq:iter-tau}
			\mathfrak{f}_{is}(\eta^{(q)}) := \bm{b}_{is}-
			\sum_n\sum_t
			\frac{ \Theta_{isnt}\eta^{(q)}_{is} }
			{\Theta_{isnt}\eta^{(q)}_{is} + \sum_{j \neq i} \Theta_{jsnt}\eta^{(q-1)}_{js}}=0
			\end{equation}
		\end{itemize}
		
		\bigskip
		{\bf If} a convergence criteria is met, {\bf Then} Stop 
	\end{enumerate}
	
\end{algorithm}
A comment on the solution of the nonlinear equations given
in~Algorithm \ref{alg:b2tau} (see \eqref{eq:iter-tau})
is in order. By a closer inspection of the
functions $\mathfrak{f}_{is}(\cdot)$ on the right side of this equation we
observe that $\mathfrak{f}^{\prime}_{is}<0$. Furthermore, the upper bounds on $\bm
b$ given in~\eqref{constraints} guarantee that $\mathfrak{f}_{is}(0)>0$ and
$\mathfrak{f}_{is}(1)<0$ whenever $\eta^{(q-1)}_{is}\in (0,1)$. Hence,
$\eta^{q}_{is}\in (0,1)$ is unique (from the monotonicity of
$\mathfrak{f}_{is}(\cdot)$) and this guarantees that $\tau^{q}_{is}$ is
also unique and positive.

\section{Discussion and conclusions}

These algorithms allow us to empirically estimate the optimal
spatial-dynamic policies to maximize the value of lakes net of damages
from invasive species. To interpret these results we focus on the
cases where $i$ and $s$ are such that $K_{is}\in [-a,a]$. In these
cases the optimal number of boating trips are defined as

\begin{equation}\label{optimalb}
	b^*_{is}=\frac{\frac{1}{|\alpha|}-\mathcal{L}^b\left(\frac{(1-x_{i(s-1)})}{2a}w_{is}+\frac{1+x_{i(s-1)}}{2}\right)
	-\mathcal{L}^hh_i\left(\frac{1-x_{i(s-1)}}{2a}\right)y_{is}A_{is}}{\mathcal{L}^b\left(\frac{1-x_{i(s-1)}}{2a}\right)y_{is}A_{is}}.
\end{equation}

We can simplify equation \eqref{optimalb} so that the optimal number of trips to each lake occurs when

\begin{equation}\label{mbmc}
	\mathcal{L}^b(x_{is}-x_{i(s-1)})+\mathcal{L}^hh_i\frac{\partial x_{is}}{\partial b_{is}} = \frac{1}{|\alpha|}.
\end{equation}

The left-hand side of equation \eqref{mbmc} is the expected marginal costs from boating. $\mathcal{L}^b(x_{is}-x_{i(s-1)})$ is the expected marginal damages to boaters from the spread of AIS and $\mathcal{L}^hh_i\frac{\partial x_{is}}{\partial b_{is}}$ is the marginal damages to homeowners from boaters spreading AIS. The right-hand side of equation \eqref{mbmc} is the marginal benefit of boating. 

Notice when the propagule pressure $A_{is}\rightarrow 0$ optimal
$b^*_{is}\rightarrow \infty$. Since we do not allow the optimal boat
ramp fee to be negative $b^*_{is}=\upsilon_{is}$ and
$\tau_{is}^*=0$. This makes intuitive sense; if AIS dispersal from
boaters is small, then the optimal policy is to have no boat ramp
fee. On the other hand when the benefits from boating are lower than
the damages from boating then the optimal number of boating trips
might reach the lower bound, which is zero\footnote{Clearly, allowing
  a negative number of boating trips would not be realistic.}. The
lower bound of boating trips corresponds to an optimal boat ramp fee
that approaches infinity, $\tau_{is}^*\rightarrow \infty$. This also
makes intuitive sense; if the damages from AIS are worse than the
benefits from boating it makes sense to close the lake to boating.

\subsection{A numerical example}
Next, we provide a simple numerical illustration of how the proposed
method works. We consider a system with $2$ lakes, $10$ boaters, $2$
seasons, and with $11$ days per season, i.e. $(I,S,N,T)=(3,1,10,11)$.
For the rest of the model parameters, we have set $\mathcal{L}^b=100$,
$\mathcal{L}^h=1400$, $\gamma=5$ , $\xi=0.1$, $f=0.03$, $h_i=1$, for
$i=1,\ldots, I$, $\lambda_{is}=-0.001$, $r=0.5\times 10^{-1}$,
$x_{\cdot,s=1}=0.5$, $\tau_{i,s}=1.0$, and $\tilde{Z}_{isnt}\sim
Uniform(-2,-1)$.  In~Table~\ref{tb:experiment} we show the optimal
number of boating trips, tax values, invaded probabilities and the
total total benefits value.  The data are for three simulations with
varying $\alpha$ in which we have the optimal value of the boating
trips $b_{is}$ to be at the upper bound, at the lower bound, or in the middle.
\begin{table}[ht]
  \caption{Optimal number of boating trips for varying values of the
    parameter $\alpha$}
  \label{tb:experiment}
  \centering
  \begin{tabular}{ | c | c |  c | c |}\hline
$\alpha$ & 0.010 & 0.133 & 0.015 \\ \hline
 upper bound (U) on $b_{i\in\{2,3\},s=2}$& 36.16, 36.39  & 53.53, 53.81 & 53.27, 53.56   \\  \hline
  lower bound (L) on $b_{i\in\{2,3\},s=2}$& 0, 0 & 0, 0 & 0, 0  \\  \hline
 optimal boating $b_{i\in\{2,3\},s=2}^*$ & 36.16, 36.39 & 34.74, 33.02 & 0.00, 0.00  \\  \hline
 optimal tax $\tau_{i\in\{2,3\},s=2}^*$ & 0.02, 0.02 & 57.64, 63.72 & 375.04, 374.71  \\  \hline
 invaded prob. $x_{i\in\{2,3\},s=2}^*$  & 0.83, 0.83 & 0.74 , 0.74  & 0.75, 0.75 \\  \hline
 $\mathcal{F}(b_{is}=L)$, $\mathcal{F}(b_{is}=U)$  & -8802, -1522 & -8054, -3678 & -1115, -3479\\  \hline
 $\mathcal{F}(b_{is}=b_{is}^*)$ & -1522 & -1669 & -1115 \\  \hline
  \end{tabular}
\end{table} 

\section{Conclusions}
In conclusion, we have developed a novel method to solve for the
optimal management of a complex spatial-dynamic process (the spread of
AIS) with an endogenous diffusion mechanism. The benefits of our
method include an analytical solution for the optimal number of trips
to each lake in each season and an iterative method to solve for the
spatial-dynamic optimal boat ramp fee $\bm{\tau}$ from $\bm{b}$ that
is likely to converge.  A careful look at the algorithms given here
show that the main computational cost is in evaluating the function
$\mathcal{F}$. Applying this novel framework to empirical data, to
other migration processes, as well as the mathematical analysis of the
proposed algorithms are subject of current and future research.

\end{document}